\documentstyle[12pt]{article}
\makeatletter

\@addtoreset{equation}{section}
\makeatother

\begin{document}
\begin{titlepage}
\begin{flushright}
DISTA-2008\\
DPUR-TH-8\\
DFPD08-TH-02\\
hep-th/yymmnnn
\end{flushright}
\vskip 1.5cm
\begin{center}
{\LARGE \bf 
Y-Formalism and Curved $\mathbf{\beta\!-\!\gamma}$ Systems
} 
\vfill
{\large Pietro Antonio Grassi$^{~1}$, Ichiro Oda$^{~2}$ and Mario Tonin$^{~3}$ } 
\\
\vfill {
$^{1}$ DISTA, Universit\`a del Piemonte Orientale, \\
via Bellini 25/g, 15100 Alessandria, Italy, and INFN - Sezione di
Torino, Italy\\
\vskip .2cm
$^{2}$
Department of Physics, Faculty of Science, \\ 
University of the Ryukyus, Nishihara, Okinawa 903-0213, Japan. \\
\vskip .2cm
$^{3}$
Dipartimento di Fisica, Universit\`a degli Studi di Padova, \\
INFN, Sezionedi Padova, Via F. Marzolo 8, 35131 Padova, Italy 
}
\end{center}
\vfill
\begin{abstract}
We adopt the $Y$-formalism to study $\beta-\gamma$ systems on 
hypersurfaces. We compute the operator product expansions of 
gauge-invariant currents and we discuss some applications of the $Y$-formalism 
to model on Calabi-Yau spaces. 
\end{abstract}
\vfill

\date{February 2008}
\end{titlepage}

\tableofcontents
\newpage
\section{Introduction} 

The technique of the $Y$-formalism developed in a series of papers 
\cite{Y-formalism:2001ab}-\cite{Y-formalism:2007ab} has been used to derive 
the complete 
operator product expansions for the composite operators involved in the 
pure spinor string theory formalism \cite{Berkovits:2000fe}. 
The $Y$-formalism is very useful to compute the contact terms and the anomaly terms in the OPE's. 
It is based on the observation that, in general to derive those terms only the local structure of the theory  is needed and not the global information on the space. Thus, for instance, given a system whose 
non-trivial information is encoded into an algebraic curve (such as $\beta-\gamma$  system 
on a hypersurface, or a topological string on a Calabi-Yau space), one can regard the system as if 
it would be free using the $Y$-formalism to impose that the fields are constrained to live on a hypersurface. Before discussing the details of the formalism and some applications to $\beta-\gamma$ system \cite{Witten:2005px}-\cite{Tan:2006by}, we would like to present some interesting models to be analyzed using the present technique. 

We start with some interesting 4d models. We consider their relations 
with complex algebras generated by the currents. 

The first model is described by four 
coordinates $\gamma^i$ (with $i=1,\dots,4$) living on the complex plane ${\bf C}^4$ (no constraint for the moment). In terms of them, one can construct 16 currents that form the $Sl(4)\oplus Gl(1)$ algebra 
with generators 
\begin{equation}
\label{algebrafour}{
J_{i}^{~j} = \beta_i \gamma^j - \frac{1}{ 4}\delta_i^{~j} J \,, ~~~~~
J = \beta_i \gamma^i\,,
}
\end{equation}
where we have introduced the conjugate momentum $\beta_i$ for each coordinate $\gamma^i$. 
The level of the algebra is easily computable by computing the double poles of the generators 
and one easily gets 
\begin{equation}\label{doublepoles}
J_{i}^{~j}(z) J_{k}^{~l}(w)  \rightarrow - \frac{\delta_i^l \delta_k^j }{ (z-w)^2} + {\rm first~order~poles}\,, ~~~~~
J(z) J(w)  \rightarrow - \frac{4 }{ (z-w)^2},
\end{equation}
and therefore we have an affine algebra of the form $Sl(4)_{-1} \oplus Gl(1)_{-4}$. 
So, the character 
\begin{equation}
\label{cfour}{
Z_{{\bf C}^4}(t_x,t_y,t_z,t_t|q) =
\frac{1}{\sqrt{t_x t_y t_z t_t}}
\frac{\eta^4(q) }{ \theta_1(t_x|q) \theta_1(t_y|q)\theta_1(t_t|q)\theta_1(t_ztq) }
},
\end{equation}
can be expanded into a sum of characters of the two algebras 
\begin{equation}
\label{deco}{
Z_{{\bf C}^4}(t_x,t_y,t_z,t_t|q)  = \sum_{K} \chi_K^{SL(4), k=-1}(t_1, t_2, t_3|q)  f_K(t)\eta_K(q)
},
\end{equation}
where the sum over $K$-representations of $SL(4)$ and $f_K(t)$ are functions of $t= (t_x t_y t_z t_t)^{1/4}$. The parameters $t_i$ are associated to the diagonal generators of $SL(4)$. 

The second example is constructed by imposing an additional constraint. 
On the set of coordinates $\gamma^i$ we impose the constraint 
\begin{equation}\label{conio}
\gamma^1 \gamma^2 - \gamma^3 \gamma^4 =0\,.
\end{equation}
This is the well-known conifold, which is a singular Calabi-Yau space and it is a non-compact 
toric variety with three dimensional complex directions. It is denoted by ${{\bf C}^4[\gamma^i]/ 
\langle \gamma^1 \gamma^2 - \gamma^3 \gamma^4=0 \rangle}$ in algebraic topology and 
as a toric variety is denoted by $C^4/C^*$.  
In this case there are only 7 currents preserving the constraint, 
by denoting $g_{ij}$ the metric in ${\bf C}^4$ with 
non-trivial entries $g_{12} = - g_{34} =1$, we have 
\begin{equation}
\label{currentsconi}{
J_{[ij]} = \beta_{[i} g_{j]k} \gamma^k\,, ~~~~~~ J = \beta_i \gamma^i\,.
}
\end{equation}
The third model is a deformation of the previous one, by adding a small deformation to the constraint 
$$
\gamma^1 \gamma^2 - \gamma^3 \gamma^4 =0 \rightarrow 
\gamma^1 \gamma^2 - \gamma^3 \gamma^4 =\epsilon \,.
$$
In this case the constraint is invariant under only the rotations of $SO(4)$ and it is not 
homogeneous so the currents $J$ do not generate a symmetry of the model. 

In the same way one can define the projective space by gauging the scale symmetry. For 
example the projective space ${\bf P}^3$ can be described in terms of the currents

\begin{equation}
\label{algebra}{
J_{i}^{~j} = w_i x^j - \frac{1}{ 4}\delta_i^{~j} J \,, ~~~~~
J = w_i x^i\,,
}
\end{equation}
where $J_i^{~j}$ generate $SL(4)$ and $J$ generates $GL(1)$. 
In the following we will discuss the details and the application of the $Y$-formalism to the case of projective 
spaces. They can be treated as gauged linear sigma models, but there are interesting anomalies 
that have to be taken into account. In addition, the projective spaces are useful to define Calabi-Yau space 
by selecting suitable sections in the space itself. 

There are also some interesting 2d models that can be considered. We can again start with the obvious case of two free coordinates
\begin{equation}
\label{fourmodels}{
{\bf C}^2 \rightarrow SL(2)_{-1} \otimes GL(1)_{-2}\,.
}
\end{equation}
Then, we can impose a simple constraint 
$$
{{\bf C}^2[x,y] \over \langle xy = 0 \rangle}   \rightarrow SO(2)_{-1} \otimes GL(1)_{+2}\,, 
$$
and finally, we can deform it to get 
$$
{\bf C}^* = GL(1)  = {{\bf C}^2[x,y] \over \langle xy =1 \rangle} \rightarrow SO(2)_{-1} \,,
$$
$$
{\bf P}^1 = \frac{{\bf C}^2 }{ {\bf C}^*} \rightarrow SL(2)_{-2}
~~~~~~{\rm charges~of}~[x,y]=(1,1).
$$
In all cases, we can use the $Y$-formalism to study the OPE's of the 
gauge-invariant operators without 
solving the constraint explicitly. 

The analysis of the $\beta-\gamma$ systems on hypersurfaces has been discussed in some papers 
until now. We would like to mention the pioneering work \cite{Malikov:1998dw}-\cite{Gorbounov:2000ac} on the construction 
of chiral vertex algbebras. These constructions already encode some of the conformal field theory 
analysis for $\beta-\gamma$ systems. However, they never explicitly compute the CFT algebra in presence of hypersurface constraints except some simple cases when the constraints are 
solvable in a simple way. Nevertheless their analysis was pivotal for more recent 
developments \cite{Witten:2005px,Nekrasov:2005wg}. 

The main issue is the quantization of a system in presence of constraints. This is a well-known 
problem in quantum field theory and it has been discussed in the literature since the 
advent of quantum mechanics and quantum field theory. Nonetheless, in the case of 2d field 
theories such as string theories and sigma models, one has the advantage of performing the 
computations in an explicit and exact way using the radial quantization technique and using the 
conformal field theory methods. In this regard, one would like to maintain such a strong feature even 
in presence of constraints. 
Therefore, one has to treat the constraints in a radically different way by imposing them at each step of computation without actually solving them. One example is the computation of the propagator  
between constrained fields. One can proceed as follows: first compute the correlator as it would have been free, then modify it consistently with the contraints. This procedure 
is encoded in the $Y$-formalism and it gives a systematical way to compute the correlation 
functions among different CFT operators. It  can be used 
to compute all possible OPE's among gauge-invariant currents (which are 
not sensible to the details of the "gauge-fixing" procedure). In addition, 
in order to give consistent results one has to check if all $Y$-dependent terms 
drop out from the computation of gauge-invariant quantities and this provides a strong check 
on the formalism itself. 

In this article, we develop the $Y$ formalism for computation of generic $\beta-\gamma$ systems. It is pointed out that 
at the present we are not able to formulate the $Y$-formalism in general and some additional ingredient might have been introduced. However, there are
a vast number of applications which can be done even by using the present status of the formalism. 

The paper is organized as follows: in sec. 2 we briefly review the curved $\beta-\gamma$ systems 
and we give some notations. In sec. 3, we develop the $Y$-formalism for $\beta-\gamma$ systemss for general systems first and then for quadratic and partly quadratic constraints. We point 
out some obstructions of the formalism. In sec. 4, we consider the gauge linear sigma model 
counterparts of some examples. In sec. 5, we discuss additional variables and we make 
the contact with the pure spinor formalism. In sec. 6, we generalize the $Y$-formalism to 
superprojective spaces. 

\section{Curved $\beta-\gamma$ systems}

A non-linear $\beta-\gamma$ system is specified by a map 
\begin{equation}
\gamma: \Sigma \rightarrow X,
\end{equation}
and a $(1, 0)$ form $\beta$ on $\Sigma$, valued in 
the pull-back $\gamma^{*}(T^{*}X)$, where $\Sigma$ is the world-sheet Riemann surface
and $X$ is a complex target space manifold. If $\lbrace U_{(\alpha)}\rbrace$ is an open covering of $X$ and $\gamma^{i}$ are local coordinates in $U_{(\alpha)}$,
the  $\beta-\gamma$ system is  described by the action 
\begin {equation}\label{action}
 S = \int \beta_{i}\bar\partial\gamma^{i},
\end{equation}
which becomes locally linear and free in $U_{(\alpha)}$.
The basic OPE is 
\begin{equation}
\beta_{i}(y)\gamma^{j}(z) = {{\delta_{i}^{\enskip j}}\over{y - z}}.
\end{equation}
The action is invariant under diffeomorphisms
\begin{equation}\label{diff0}
\delta \gamma^i = V^i(\gamma)\,, ~~~~
\delta \beta_i = - \beta_j \partial_i V^j(\gamma)\,.  
\end{equation}
where $V = V^{i}\partial_{i}$ is a holomorphic vector field on $X$ and we use 
the notation
$\partial_{i} = {{\partial} \over {\partial \gamma^{i}}}$. 
The corresponding current is 
\begin{equation}\label{J}
J_{V} = \beta_{i}V^{i}.
\end{equation}
Given two holomorphic vector fields $V$ and $U$, we can compute the OPE
\begin{equation}
J_{V}(y) J_{U}(z) =  - {{\partial_{i}V^{j}\partial_{j}U^{i}(z)}\over{(y-z)^{2}}} 
+ {{J_{[V,U]}}\over{y-z}} - {{(\partial_{k}\partial_{i}V^{j})
(\partial_{j}U^{i})}\over{y-z}} \partial \gamma^k.
\end{equation}
The last term shows a failure of closure for diffeomorphisms and it is a 
possible source of an anomaly that arises  if this term cannot be reabsorbed 
by a  redefinition of the 
currents. As  shown in \cite{Witten:2005px, Nekrasov:2005wg}, this obstruction arises if the 
first Pontryagin class $p_{1}(X)$ of $X$ does not vanish. In fact the
diffeomorphisms must be used to glue together the different patches $U_
{(\alpha)} $ of $X$ and the anomaly, which arises if $p_{1}(X)\ne 0$, is 
indeed an obstruction for $X$ to be defined globally.
Another anomaly is present \cite{Witten:2005px, Nekrasov:2005wg} if the product $ c_{1}(\Sigma)c_
{1}(X)$ of the first Chern classes of $\Sigma$ and $X$ does not vanish.
However, this anomaly does not appear if one works on a world-sheet with
 $ c_{1}(\Sigma) = 0 $. 
Having done these dutiful specifications, we will be no more concerned with   
these obstructions henceforth.\par

In a vast class of interesting  models  the target space 
manifold $X$ is a hypersurface in $n$ dimensions defined by one or more 
constraints. A well-known example is that of pure spinors in $D=10$, where 
the manifold is a cone over  $ SO(10)/U(5)$ \cite{Berkovits:2005hy}.
In this paper we will treat mainly the case of only one 
constraint
\begin{equation} \label{constraint}
\Phi(\gamma) = \sum_{h=0}^{n}{ \Phi^{(h)}(\gamma)} =0\,,
\end{equation}
with $\Phi^{(h)}$ being a homogeneous function of degree $h$.
In presence of the constraint (\ref{constraint}), the action
(\ref{action}) is invariant under the gauge symmetry 
\begin{equation}\label{gaugesymm}
\delta \beta_i = \lambda \partial_{i} \Phi(\gamma)\,.
\end{equation} 
Notice that the gauge parameter $\lambda $ is a $(1, 0)$ form.
This setting  is 
useful to study the example of hypersurfaces in projective spaces.
The constraint $\Phi(\gamma)=0$ to be embedded into a projective space needs 
to be homogeneous of some degree $h > 1$ with respect to the rescaling of the 
coordinates so that only a term survives in (\ref{constraint}) and not only 
the action but also the constraint are invariant under the scale transformation 
$\gamma^i \rightarrow \Lambda \gamma^i$ where $\Lambda$ is a scale. 
If this symmetry is gauged, so that $\Lambda$ becomes local, the model describes a projective space. 
Therefore, for a hypersurface in a projective space there are two sets of 
gauge symmetries and of constraints. The first set is composed by a 
scaling gauge symmetry and a linear constraint $\beta_
{i}\gamma^{i} = 0 $.
The second one is determined by the constraint 
$\Phi(\gamma)=0$ which is non-linear and by the gauge symmetry 
(\ref{gaugesymm}) which is non-linear.

\section{$Y$-formalism for constrained $\beta-\gamma$ systems}
\subsection{$Y$-formalism}

The usual approach to compute OPE's for a constrained $\beta-\gamma$ 
model consists in solving first the constraints in a given chart where the
model becomes free, computing the free OPE's for this reduced system and then
reconstructing the OPE's for the original operators. 

The $Y$-formalism avoids this procedure by postulating the basic OPE among 
$\beta_{i}$ and $\gamma^{j}$ in the form
\begin{equation} \label{basic}
< \beta_{i}(y)\gamma^{j}(z)> = {{1}\over{y-z}} (\delta_{i}^{\enskip j} - 
K_{i}^{\enskip j}(z)),
\end{equation}
and choosing $K_{i}^{\enskip j} $ by demanding that the OPE's of $\beta_i$
(and therefore of any operator) with the constraints vanish. 
Note that $K_{i}^{\enskip j} $ depends on a non-covariant field $Y_i$. 
For instance, 
in the case of pure spinors 
where the constraint is $\lambda \Gamma^{m} \lambda = 0$, the basic OPE is
\begin{equation}
<\omega_{\alpha}(y)\lambda^{\beta}(z)> = {{1}\over{y-z}} 
(\delta_{\alpha}^{\enskip \beta} - \frac{1}{2} (\Gamma_{m}\lambda)_{\alpha}
(Y\Gamma^{m})^{\beta}(z)),
\end{equation}
with $Y_{\alpha} \equiv {v_{\alpha} \over (v\lambda)}$, $v_{\alpha}$ being a 
constant spinor. This basic OPE was proposed in \cite{Berkovits:2000fe} 
for the first
time. If one makes use of it to compute OPE's among gauge-invariant operators,
 one obtains poles with $Y$-dependent contributions. 
However, it has been shown in \cite{Y-formalism:2005ac} that if one introduces
 suitable $Y$-dependent corrections in the definition of these operators, 
one can obtain $Y$-independent OPE's and they coincide with
the OPE's obtained by the usual method mentioned at the beginning of this 
section.

In this section we will extend the $Y$-formalism to $\beta-\gamma$ system
described by the action (\ref{action}) on a hypersurface defined by the
constraint (\ref{constraint}). In particular, we shall discuss when this
formalism is applicable and what its limitations are.

Let us consider a constrained $\beta-\gamma$ model
\begin{equation}\label{action0}
S = \int \beta_{i}\bar\partial\gamma^{i},
\end{equation}
where the index $i$ runs over $1, \cdots, N$ and this system is
characterized  by the constraint 
\begin{equation} \label{constraint2}
\Phi(\gamma) = \sum_{h=0}^{n}{ \Phi^{(h)}(\gamma)} = 0. 
\end{equation}
Here 
\begin{equation}
\Phi^{(h)}(\gamma) = {{1}\over{h}} g_{i_{1} \cdots i_{h}}\gamma^{i_{1}} 
\cdots \gamma^{i_{h}},
\end{equation}
is homogeneous in $\gamma$ of degree $h$.  
Let us introduce the following notation: 
\begin{equation}
\Phi_{i_{1} \cdots i_{p}}(\gamma) =
\partial_ {i_{1}} \cdots \partial_{i_p} \Phi(\gamma),
\end{equation}
with the definition of 
$\partial_{i} \equiv {{\partial}\over{\partial \gamma^{i}}}$.
The action (\ref{action0}) has a local symmetry 
\begin{equation}
\delta \beta_{i} = \lambda \Phi_{i},
\end{equation}
where $\lambda$ is a local gauge parameter.
The classically gauge-invariant operators are 
\begin{equation}
T^0  = \beta_i \partial \gamma^i,
\end{equation}
and
\begin{equation}
J^0_{rs} = \beta_{[r} \Phi_{s]}.
\end{equation}
These currents are conserved since they leave invariant the constraint and 
the action. 

The other gauge-invariant and conserved currents can exist depending on 
the form of $\Phi(\gamma)$.  For instance let us consider a group $G$ of rigid transformations: 
\begin{equation}
\delta \gamma^i = \delta\Lambda^I (P_I)^i_{\enskip j}\gamma^{j}\,, ~~~~
\delta \beta_i = - \delta\Lambda^I \beta_j (P_I)^j_{\enskip i}\,,  
\end{equation}
where $\delta\Lambda_{I}$ are constant (infinitesimal) parameters with
the index $I$ running over the rank of $G$ and $P_{I}$
is a representation of the Lie algebra ${\cal G}$ of $ G $: 
$$ \Big[P_{[I},P_{J]}\Big] = f_{[I J]}
^{K}P_{K}$$ where $f_{[I J]}^{K}$ are the structure
 constants of $ {\cal G}$ and suppose that $\Phi(\gamma)$ is scalar, that is 
\begin{equation}
\Phi_{i}(P_{I})^{i}_{\enskip j}\gamma^{j} = 0
\end{equation}
Then the classical $G$-current 
\begin{eqnarray}
J^{0}_{I} =\beta_{i}(P_{I})^{i}_{\enskip j} \gamma^{j},
\label{14}
\end{eqnarray}
is conserved and gauge invariant.
Also, one can have a "ghost" current if it is possible 
to assign  a ghost number $ g^{(i)}$ 
to each $\gamma^{i}$  in such a way
that $\Phi$ has ghost number $ 2 g^{(0)} $
\begin{eqnarray}
\sum g^{(i)}\gamma^{i} \Phi_{i} = 2 g^{(0)}\Phi(\gamma).
\label{15}
\end{eqnarray}
Then the classical current 
\begin{eqnarray}
J^0 = \sum g^{(i)} \beta_i \gamma^i,
\label{16}
\end{eqnarray}
becomes gauge invariant and conserved. Furthermore, if $\Phi(\gamma)$ is homogeneous in $\gamma$, the ghost current reads
\begin{eqnarray}
J^0 =  \beta_i \gamma^i.
\label{17}
\end{eqnarray}
In order to define the basic OPE (\ref{basic}), we choose a constant vector
$v^i$ and define 
\begin{eqnarray}
Y^i = \frac{v^i}{(v^j \Phi_j)},
\label{18}
\end{eqnarray}
{}from which we have a relation $Y^i \Phi_i = 1$.
At this stage, the basic OPE is given by
\begin{equation} \label{basic2}
<\beta_{i}(y) \gamma^{j}(z)> = {{1}\over{y-z}}( \delta_{i}^{\enskip j} -
\Phi_{i}(z)Y^{j}(z)) \equiv {{1}\over{y-z}}( \delta_{i}^{\enskip j} - 
K_{i}^{\enskip j}(z)). 
\end{equation}
One can then check immediately that 
\begin{eqnarray}
<\beta_{i}(y)\Phi(\gamma(z))> = 0.
\label{19}
\end{eqnarray}
Now the problem is to understand in which cases this formalism is consistent,
in other words, when it is possible to add $Y$-dependent terms to the relevant
(gauge-invariant) operators in such a way that the corresponding OPE's are
free of $Y$-dependent contributions. 

\subsection{$\beta-\gamma$ models with quadratic constraint}

Before discussing the general case, let us consider a simpler case: a class of 
models with the quadratic constraint 
\begin{equation}
\Phi(\gamma) = \frac{1}{2} \gamma^{i}g_{ij}\gamma^{j}= 0,  
\end{equation}
 where $g_{ij}$ is a constant, invertible, $N \times N$ matrix with inverse
 $g^{ij}$. The matrices $g_{ij}$ and $g^{ij}$ can be used for raising and 
lowering indices, for instance, $\gamma_{i} = g_{ij}\gamma^{j}$.
Then the basic OPE is
\begin{equation}\label{basic3}
< \beta_{i}(y)\gamma^{j}(z)> = {{1}\over{y-z}} (\delta_{i}^{\enskip j} - 
K_{i}^{\enskip j}(z)),
\end{equation}
where 
\begin{equation}
K_{i}^{\enskip j} = \gamma_{i} Y^{j},  
\end{equation}
thereby we can prove $<\beta_{i}(y) (\gamma^{j}g_{jk}\gamma^{k})(z)> = 0$.

These models have the local symmetry 
\begin{equation}
\delta \beta_i = \lambda \gamma_i,  
\end{equation}
and the classically gauge-invariant composite fields are the currents
$J^0= \beta_{i}\gamma^{i}$, $ J^0_{rs} = \beta_{[r}\gamma_{s]}$ and
the stress-energy tensor $ T^0 = \beta_{i}\partial \gamma^{i}$.\par

We shall utilize the $Y$-formalism to compute the OPE's for the corresponding 
quantum operators $J$, $J_{rs}$ and $T$, which are corrected by $Y$-dependent terms in order to have $Y$-independent OPE's.
By computing the OPE between $J^0_{rs}$ and $J^0_{pq}$ one obtains
\begin{eqnarray}
< J^0_{rs}(y) J^0_{pq}(z)> &=& - {{1}\over{(y-z)^{2}}}[ g_{sp}g_{rq} 
- K_{ps}(y)g_{qr} - K_{rq}(z)g_{sp} + K_{ps}(y)K_{rq}(z)]
\nonumber\\
&-& {{1}\over{y-z}} [\beta_{r} g_{ps}\gamma_{q} - \beta_{p}g_{rq}\gamma_{s}], 
\label{ope1}
\end{eqnarray}
(here antisymmetrization among $r, s$ and $p, q$ is implicitly understood)
which contains spurious $Y$-dependent poles. However, it can be verified by 
computing the OPE's of $J^0_{rs}$ with $Y_{[r}\partial \gamma_{s]}$ and
$ \partial Y_{[r}\gamma_{s]}$ that if one defines
\begin{eqnarray}\label{current}
J_{rs} = J^{0}_{rs} - Y_{[r}\partial \gamma_{s]} - {{1}\over{2}} \partial 
Y_{[r} \gamma_{s]},
\end{eqnarray}
the spurious terms are precisely canceled and one gets the $Y$-independent OPE
\begin{eqnarray}\label{opecurrent}
<J_{rs}(y)J_{pq}(z)> = {{1}\over{(y-z)^{2}}}g_{r[p}g_{q]s} + {{1}\over{y-z}}
(g_{p[r}J_{s]q} - g_{r[p}J_{q]s}).
\end{eqnarray}
In a similar way, provided that one defines 
\begin{eqnarray}\label{currentbis}
J = J^{0} - {{3}\over{2}} \partial Y_{i}\gamma^{i},
\end{eqnarray}
and 
\begin{eqnarray} \label{currentter}
T = T^{0} + {{1}\over{2}} \partial (Y_{i}\partial \gamma^{i}),
\end{eqnarray}
the spurious $Y$-dependent poles  also cancel in the OPE's of $J$ and $T$ with $J_{rs}$. Moreover, it turns out that the remaining OPE's of $J$ and $T$ among them are also free of spurious, $Y$-dependent contributions. 

To summarize, the algebra among currents in addition to (\ref{opecurrent})
 takes the form
\begin{eqnarray}
<J(y) J_{rs}(z)> = 0,
\label{qcurrent1}
\end{eqnarray}
\begin{eqnarray}
<T(y)J_{rs}(z)> = {{1}\over{(y-z)^2}}J_{rs}(y),
\label{qcurrent2}
\end{eqnarray}   
\begin{eqnarray}
<T(y) T(z)> =  {{N-1}\over{(y-z)^{4}}} +  {{1}\over{(y-z)}^{2}}(T(y) + T(z)),
\label{qcurrent3}
\end{eqnarray}
\begin{eqnarray}
<T(y) J(z)> =  {{N-2}\over{(y-z)^{3}}} + {{1}\over{(y-z)^{2}}}J(y),
\label{qcurrent4}
\end{eqnarray}
\begin{eqnarray}
<J(y) J(z)> =  {{4 - N}\over{(y-z)^{2}}}.
\label{qcurrent5}
\end{eqnarray}
One should notice that in the OPE's $<J J>$ and $<T J>$, there are $Y$-independent contributions, given by $ {3 \over (y-z)^2} $ and ${{-1} \over {(y-z)^{3}}}$ respectively, coming from the $Y$-dependent terms in the definition of $J$ and $T$.

\subsection{An obstruction for models with constraints of degree greater 
than two}

Now let us consider the general case where the constraint (\ref{constraint2})
is a polynomial in $\gamma^i$ of order greater than two. We shall show that in 
this case the $Y$-formalism does not work in general. To see why, let us compute
the OPE $<J_{rs}J_{pq}>$.
The most general expression for $J_{rs}$ is
\begin{eqnarray} \label{current2}
J_{rs} = J^{0}_{rs} -c_{1} Y_{[r}\partial \Phi_{s]} - c_{2} \partial 
Y_{[r} \Phi_{s]},
\end{eqnarray}
where $c_{1}$ and $c_{2}$ are constants. Here $\Phi_{ij}$ are used for lowering indices such as $Y_i = Y^j \Phi_{ji}$. But the problem arises because the OPE's of 
$J^0_{rs}$  with the $Y$-dependent terms $Y_{[i}\partial \Phi_{j]}$ and 
$\partial Y_{[i} \Phi_{j]}$ yield poles with residuum proportional to 
$Y^{k} \Phi_{k i p}$ which cannot 
be canceled. 
Indeed, we can easily derive 
\begin{eqnarray}
<J^0_{rs}(y)Y_{i}(z)> =  -{{1}\over{y-z}}
Y_{[r} K_{s]i} + {1 \over{y-z}} Y^{j} \Phi_{[s} \Phi_{r]ji}.
\end{eqnarray}
As before, for instance, if one selects $c_{1} = 1$ and 
$c_{2} = {{1} \over {2}}$ in order to cancel the $Y$-dependent double 
poles, one has the equation
\begin{eqnarray}
&{}& <J_{rs}(y)J_{pq}(z)> 
= {{-1} \over{(y-z)^2}} [ \Phi_{ps}(y) \Phi_{qr}(z) + ( \Phi_s \Phi_q Y^i
\Phi_{ipr} )(z) ]
+  {1 \over{y-z}}[ \Phi_{qr} J_{ps} - \Phi_{ps}J_{rq} ]
\nonumber\\
&+& {1 \over{y-z}} [ - Y^i \Phi_{ipr} \Phi_s \partial \Phi_q +  Y^{i}Y^{j}
\partial \gamma^{k}(\Phi_{ir}\Phi_{jkp} - \Phi_{ip}\Phi_{jkr})
- \frac{1}{2} \partial (Y^i \Phi_{ipr}) \Phi_s \Phi_q], 
\label{ope3}
\end{eqnarray}
which is obviously inconsistent unless 
\begin{eqnarray}
Y^{i} \Phi_{ijk} = 0.
\end{eqnarray}
(In (\ref{ope3}), antisymmetrization among $r, s$ and $p, q$ is understood.)

One could think that the problem might depend on our choice of $Y^{i}$ and a 
different choice would avoid the problem, but it is not so.
The most general possibility is to start with a constant vector $v_{i}$ and 
to define $Y^{i}$ as $ Y^{i}= {{v_{j} A^{j i}(\gamma)} \over {(v_{j} 
A^{j i}(\gamma)\Phi_{i})}}$. If $\Phi_{i j}$ is invertible, calling 
${\tilde \Phi}^{i j}$ its inverse, one can choose $ A^{ij} = {\tilde \Phi}^{ij}$ so that 
$Y_{i} = {{v_{i}} \over {(v_{j}{\tilde \Phi}^{ji}\Phi_{i})}} $. However, 
the problem remains in general since in this case  
\begin{eqnarray}
<J^0_{rs}(y) Y_i(z)> =  
-{{1}\over{y-z}} Y_{[r} K _{s] i} 
+ {1 \over{y-z}} Y_i Y^j \Phi_{[s} \Phi_{r]j k} {\tilde \Phi}^{k l}\Phi_l.
\label{A1}
\end{eqnarray}

A possible exception is the case where the constraint is homogeneous of degree
$h$, for which we have following identities:
\begin{eqnarray}
\Phi_{i}\gamma^i &=& h \Phi, \nonumber\\
\Phi_{i j}\gamma^j &=& (h - 1) \Phi_{i}, \nonumber\\
\Phi_{i j k}\gamma^{k} &=& (h - 2) \Phi_{i j}.
\end{eqnarray}
Then, we have relations
\begin{eqnarray}
\gamma^j &=& (h-1) \tilde\Phi^{ji}\Phi_{i} , \nonumber\\
Y_{i} &=& (h-1) {{v_{i}}\over {(v_{j}\gamma^{j})}}.
\label{A2}
\end{eqnarray}
Thus, using Eq's. (\ref{A1})-(\ref{A2}), one finds
\begin{eqnarray}
<J^0_{rs}(y) Y_{i}(z)> =  -{{1}\over{y-z}}{{1}\over{h-1}}
Y_{[r} K_{s]i}.
\end{eqnarray}
Note that the OPE $<J^0_{rs} J^0_{pq}>$ is given by 
\begin{eqnarray}
<J^0_{rs}(y) J^0_{pq}(z)> &=& - \frac{1}{(y-z)^2} [ \Phi_{rq}(z)
\Phi_{ps}(y) - \Phi_{rq}(z) K_{ps}(y) - K_{rq}(z) \Phi_{ps}(y)
+ K_{rq}(z) K_{ps}(y) ]
\nonumber\\
&-& \frac{1}{y-z}(\beta_r \Phi_q \Phi_{ps} - \beta_p \Phi_s \Phi_{rq}).
\end{eqnarray}
On the other hand, one can calculate
\begin{eqnarray}
&{}& <J^0_{rs}(y) (\partial\Phi_{[p} Y_{q]})(z)> 
+ < (\partial\Phi_{[r}Y_{s]})(y)  J^0_{pq}(z) > 
= {{-1}\over{(y-z)^{2}}}[ \Phi_{rq}(z) K_{ps}(y) + \Phi_{ps}(y) K_{rq}(z)] 
\nonumber\\
&+& {{1}\over{y-z}} [\Phi_{rq} \partial \Phi_p Y_s - 
\Phi_{ps} \partial \Phi_r Y_q] 
+ {{1}\over{y-z}} [K_{rq} \partial \Phi_p Y_s
- K_{ps} \partial \Phi_r Y_q ] {{1}\over{h-1}},
\end{eqnarray}
and 
\begin{eqnarray}
&{}& < J^0_{rs}(y)(\Phi_{[p}\partial Y_{q]})(z)> + < (\Phi_{[r} \partial
Y_{s]})(y)  J^0_{pq}(z) > 
= {{2}\over{(y-z)^{2}}} K_{rq}(z) K_{ps}(y) {{1}\over{h-1}}
\nonumber\\
&-& {{2}\over{y-z}} [ K_{rq} \partial \Phi_p Y_s 
- K_{ps} \partial \Phi_r Y_q ] {{1}\over{h-1}}
+ {{1}\over{y-z}} [ \Phi_{qr} \Phi_p \partial Y_s - 
\Phi_{ps} \Phi_r \partial Y_q ].
\end{eqnarray}
(In the above, antisymmetrization among $r, s$ and $p, q$ is implicitly understood again.)
Taking $c_{1} = 1$ and $c_{2} = {{1} \over {2}} (h-1)$ in (\ref{current2}) 
in order to cancel the $Y$-dependent terms in the double pole, one gets
\begin{eqnarray}
<J_{rs}(y)J_{pq}(z)> &=& {{-1}\over{(y-z)^{2}}} \Phi_{ps}(y) \Phi_{qr}(z) +
{{1}\over{y-z}}( \Phi_{qr}J_{ps} - \Phi_{ps}J_{rq})
\nonumber\\
&+& \frac{1}{y-z} \frac{2-h}{h-1} [ K_{rq} \partial \Phi_p Y_s 
- K_{ps} \partial \Phi_r Y_q ].
\end{eqnarray}
This result is consistent only if $h=2$, namely, if the constraint is 
quadratic.  

\subsection{Models with partly quadratic constraint}

It can happen that for special models and with suitable choice
of $Y^i$, the problem pointed out in the previous subsection 
is absent even if the constraint $\Phi$ is more than quadratic in $\gamma^i$.

Indeed, suppose that the fields $\gamma^{i}$ can be splitted in two sets
\begin{eqnarray}
\gamma^i = ( \gamma^{a} \equiv y^{a}, \gamma^{\hat a}\equiv x^{\hat a}),
\end{eqnarray}
where $a = 1, \cdots, n_1$, $\hat a = 1, \cdots, n_2$ ($n_{1} + n_{2} = N$), and that the constraint can be written as 
\begin{equation} \label {newconstraint}
\Phi(y,x) \equiv \Phi^{(1)}(y) + \Phi^{(2)}(x) = 0, 
\end{equation}
where 
\begin{eqnarray}
\Phi^{(1)}(y) = \frac{1}{2} y^a g_{ab} y^b,
\end{eqnarray}
is quadratic in $y^{a}$ while $\Phi^{(2)}(x)$ is a generic polynomial in  $x^{\hat a}$.
The $Y$-formalism can be adapted to this class of models by choosing 
$Y^{i}$ (that is, $v^{i}$) to have non-vanishing components only in 
the direction of $y^a$
\begin{eqnarray}
Y^i = Y^a \delta_a^i.
\end{eqnarray}
With this choice, the formalism turns out to be consistent. In fact, in this case, the condition $Y^{i} \Phi_{ijk} = 0$ 
is satisfied. 
The action is
\begin{eqnarray}
S = \int \beta_{a}\bar\partial y^{a} 
+ \int \beta_{\hat b}\bar\partial x^{\hat b},
\end{eqnarray}
and the model has the local symmetry 
\begin{eqnarray}\label{gaugeA}
\delta \beta_i = \lambda \Phi_i(x, y).
\end{eqnarray}
The classically gauge-invariant operators are 
\begin{eqnarray}
T^0  &=& \beta_i \partial \gamma^i, \nonumber\\
J^0_{ij} &=& \beta_{[i} \Phi_{j]}.
\end{eqnarray}
If the constraint $\Phi(x, y)$ is generic, $J_{i j}$ and $T$ are 
the only gauge-invariant operators, but there are further 
gauge-invariant operators for a specific form of constraints.

In order to observe this fact, let us suppose that one can assign a 
"ghost number" $g^{(a)}$ and $g^{(\hat b)}$ to $y^{a}$  and $x^{\hat b}$, respectively,  in such a way that both $\Phi^{(1)}
(y)$ and $\Phi^{(2)}(x)$ have ghost number $2g^{(0)}$.
Then, the classical "ghost current"
\begin{eqnarray}
J^0 = \sum g^{(a)}\beta_{a}y^{a} + \sum g^{(\hat b)}\beta_{\hat b} 
x^{\hat b},
\end{eqnarray}
is gauge invariant and conserved.

Moreover, if $y^a$ and $x^{\hat b}$ belong to two representations
of a Lie group $G$
\begin{eqnarray}
\delta y^{a} = \delta \Lambda^{I} (P_{(1)I})^{a}_{\enskip b}y^{b},
\qquad \delta x^{\hat a} = \delta\Lambda^{I} 
(P_{(2)I})^{\hat a}_{\enskip\hat b}x^{\hat b},
\end{eqnarray}
and $\Phi^{(1)}(y) $ and $ \Phi^{(2)}(x)$ are scalars, the classical
$G$-current 
\begin{eqnarray}
J^0_{I} = \beta_{b}( P_{(1)I})^{b}_{\enskip 
a} y^{a}  + \beta_{\hat b}( P_{(2) I})^ {\hat b}_{\enskip \hat a} x^{\hat a},
\end{eqnarray}
is also gauge invariant and conserved. 
Notice that with the constraint (\ref{newconstraint}) the currents 
$J^0_{(1) I} = \beta_{b}( P_{(1)I})^{b}_{\enskip a} y^{a}$ and $J^0_{(2) I} 
= \beta_{\hat b}( P_{(2) I})^{\hat b}_{\enskip \hat a} x^{\hat a} $ are separately gauge invariant and conserved.

It is always possible to redefine $g^{(a)}$ as $ g^{(a)} = g^{(0)} + q^{(a)}$
 so that $\sum q^{(a)}y^{a}g_{ab}y^{b} = 0 $ ; then the ghost current becomes
\begin{eqnarray} \label{ghostcurrent2}
J^0 = g^{(0)} \sum \beta_{a}y^{a} + \sum g^{(\hat b)}\beta_{\hat b} 
x^{\hat b}.
\end{eqnarray}
Moreover, if it is possible to assign charges $q^{(\hat a)}$ to $x^{\hat a}$
such that $\sum q^{(\hat a)} x^{\hat a}\Phi^{(2)}_{\hat a} = 0 $, 
$\hat J^0 = \sum q^{(a)}\beta_{a}y^{a} + \sum q^{(\hat b)}\beta_{\hat b}
 x^{\hat b}$ is gauge invariant and conserved.
$\hat J^0$ is a particular $G$-current with $G= U(1)$. Notice that if 
$\Phi^{(2)}$ is homogeneous of degree $h > 0$, one can write $  g^{(\hat b)} = {{2}\over{h}}
g^{(0)} + q^{(\hat b)}$ and the ghost current (rescaled to have $g^{(0)}= 1$)  
becomes
\begin{eqnarray}
J^0 = \sum \beta_{a}y^{a} +{{2}\over{h}} \sum \beta_{\hat b} x^{\hat b}.
\end{eqnarray}

Now we want to determine the $Y$-dependent correction terms in these currents
such that the spurious $Y$-dependent terms do not appear in the OPE's among
these currents. In addition, we wish to compute these OPE's.

In this case, the basic OPE 
\begin{eqnarray}
<\beta_{i}(y)\gamma^{j}(z)> = {{1}\over{y-z}}(\delta_{i}^{
\enskip j} - \Phi_{i}(z)Y^{j}(z)),
\end{eqnarray}
splits as follows:
\begin{eqnarray}
<\beta_{a}(y) y^{b}(z)> &=& {{1}\over{y-z}}(\delta_{a}^{\enskip b} - 
y_{a}(z)Y^{b}(z)),
\nonumber\\
<\beta_{a}(y) x^{\hat b}(z)> &=& 0,
\nonumber\\
<\beta_{\hat a}(y) x^{\hat b}(z)> &=& {{1}\over{y-z}}
\delta_{\hat a}^{\enskip \hat b},
\nonumber\\
<\beta_{\hat a}(y) y^{b}(z)> &=& - {{1}\over{y-z}}
\Phi_{\hat a}(z) Y^{b}(z),
\end{eqnarray}
where we have defined as $y_{a} = g_{ab} y^{b}$.
As a preliminary, let us coinsider the OPE's  
\begin{eqnarray}
<\beta_j(y) Y^k(z)> &=& - {{1}\over{y-z}} [ Y_j Y^k 
- \Phi_j (Y^l Y_l) Y^k ],
\nonumber\\
<\beta_j(y) Y_k(z)> &=& - {{1}\over{y-z}} [ Y_j Y_k 
- \Phi_j (Y^l Y_l) Y_k].
\end{eqnarray}
The terms proportional to $\Phi_j (Y^l Y_l)$ do not contribute 
to the OPE's of $Y$ and $\partial Y$ with the currents $J_{ij}$, $T$ and the 
other currents when they exist. In fact, in the OPE of $\beta_{[i}
\Phi_{j]}$ with $Y$ and $\partial Y$, they give rise to terms 
proportional to $\Phi_{[i} \Phi_{j]}$ and  $ \partial(\Phi_{[i}
\Phi_{j]})$.
In the OPE of $\sum g^{(i)} \beta_i \gamma^i$
with $Y$ and $\partial Y$, they give rise to terms proportional to 
$\sum g^{(i)} \Phi_i \gamma^i = 2 g^{(0)} \Phi = 0$ and to 
$\partial\Phi = 0$. In the OPE of $\beta_{i} \partial\gamma^{i}$, we
have terms proportional to $\Phi_i \partial \gamma^i = \partial \Phi = 0$ 
and to $\partial^2 \Phi = 0$ and so on.

Moreover, let us notice that 
\begin{eqnarray}
< \sum (g^{(i)} \beta_i \gamma^i)(y) \Phi_j(z)> 
&=&  {{1}\over{y-z}} \sum g^{(i)} \gamma^i \Phi_{ji}
\nonumber\\
&=&  {{1}\over{y-z}} [\partial_{j} (\sum g^{(i)} \gamma^i \Phi_i) 
- \sum g^{(i)} \delta_j^i \Phi_i] 
\nonumber\\
&=&  {{1}\over{y-z}}(2 g^{(0)} - g^{(j)}) \Phi_j.
\end{eqnarray}
Given that, the remaining calculations are straightforward. It then turns out that, in agreement with (\ref{current}) and (\ref{currentter}), the $Y$-dependent corrections for $J_{i j}$ and $T$ are given by 
\begin{eqnarray}
J_{ij} &=& J^0_{ij} - Y_{[i}\partial \gamma_{j]} - {{1}\over{2}} \partial 
Y_{[i} \gamma_{j]},
\nonumber\\
T &=& T^0 + \frac{1}{2} \sum_a \partial (Y_a \partial y^a).
\end{eqnarray}
The OPE's among these operators are
\begin{equation}
 <J_{ij}(y)J_{kl}(z)> 
= {{1} \over{2(y-z)^2}} [ \Phi_{k[i}(y) \Phi_{j]l}(z)
- \Phi_{l[i}(y) \Phi_{j]k}(z) ]
+  {1 \over{y-z}}[ \Phi_{k[i} J_{j]l} - \Phi_{l[ i}J_{j]k} ],
\end{equation}
\begin{eqnarray}
<T(y) J_{ij}(z)> =  {{1}\over{(y-z)^{2}}} J_{ij}(y),
\end{eqnarray}
\begin{eqnarray}
<T(y) T(z)> =  {{N - 1}\over{(y-z)^4}} +  {{1}\over{(y-z)^2}}(T(y) + T(z)).
\end{eqnarray}

As for the current $J$, when it exists, one gets
\begin{equation}
 J = J^0 +3/2 g^{(0)} \sum Y_{a}\partial y^{a},
\end{equation}
where $J^0$ is given in (\ref{ghostcurrent2}).
Then the relevant OPE's read
\begin{eqnarray}
<J(y) J_{i j}(z)> =  {{1/2}\over{(y-z)^{2}}} (g^{(j)} - g^{(i)})\Phi_{i j}
+ (2g^{(0)}-g^{(i)} - g^{(j)}){{1}\over{y-z}} J_{i j},
\end{eqnarray}
\begin{eqnarray}
<T(y) J(z)> = {{1}\over{(y-z)^3}} (\sum g^{(i)} - 2 g^{(0)}) +
 {{1}\over{(y-z)^2}}J(y),
\end{eqnarray}
and
\begin{eqnarray}
<J(y) J(z)> =  {{1}\over{(y-z)^{2}}} (4 (g^{(0)})^{2} - \sum g^{( i)}g^{(i)}).
\end{eqnarray}
If the constraint allows for a $G$-current $J^0_{I} = \beta_{i}(P_{I})^{i}_{
\enskip j}\gamma^{j}$ one finds that the corrected form of this current is
\begin{eqnarray}
J_{I} = (\beta P_{I} \gamma) - (Y P_{I} \partial \gamma) - 1/2 (\partial Y P_{I}\gamma),
\end{eqnarray}
and the relevant OPE's are
\begin{equation}
< J_{I}(y) J_{K}(z)> = - {{1}\over{(y-z)^2}} Tr(P_I P_K) + {{1}\over{y-z}} f_{KI}^{\enskip J}J_{J}, 
\label{J-algebra2}
\end{equation}
\begin{equation}
< J_{I}(y) J_{i j}(z)> =  {{1}\over{2 (y-z)^2}} 
[\Phi_{il}(P_{I})^l_{\enskip j} 
- \Phi_{jl}(P_{I})^l_{\enskip i}]
+ {{1}\over{y-z}}[ (P_{I})_{i}^{\enskip l}J_{l j} -
 (P_{I})_{j}^{\enskip l}J_{l i} ],
\end{equation}
\begin{equation}
<T(y) J_{I}(z)> =  {{1}\over{(y-z)^3}}Tr(P_{I}) +  {{1}\over{(y-z)^2}} 
J_{I}(y),
\end{equation}
when $ Tr(P_{I})$ is different from zero only if $G$ contains $U(1)$-factors.

\section{Examples}
\subsection{Projective Spaces}

An interesting class of $\beta-\gamma$ systems, either constrained or 
unconstrained, are the gauged $\beta-\gamma$ models. They appear for two 
reasons: the first one is the construction of model defined on a space whose 
description is given in terms of some free coordinates modulo some gauge symmetry. For 
example the projective space which can be formulated as quotients of flat $n$ dimensional 
complex spaces modulo some toric gauge symmetries. The second reason is that in some 
case hypersurfaces can be viewed as manifolds with some gauge symmetries. 
Obviously, this relation has been used in literature 
for many important results (see for example \cite{Hori:2003ic}), but it has not been developed for 
$\beta-\gamma$ systems (some considerations can be found in \cite{Tan:2006zg}). The $Y$-formalism can be also used 
in the present context, where first one uses the gauge symmetry to remove unwanted coordinates and 
then uses the $Y$-formalism to compute the remaining OPE's.  

Let us consider a group $G$ of rigid transformations: 
\begin{equation}
\delta \gamma^i = \delta\Lambda^I (P_I)^i_{\enskip j}\gamma^{j}\,, ~~~~
\delta \beta_i = - \delta\Lambda^I \beta_j (P_I)^j_{\enskip i}\,,  
\end{equation}
under which, not only the action trivially, but also the constraint, if it 
exists, is invariant.

We can  promote the constant parameters $\delta\Lambda_I$ to local ones,  by  
adding to the action a set of gauge fields $\bar A^I$ coupled to the currents 
(constraints)
$J_{I} = \beta_{i} (P_{I})^{i}_{\enskip j}\gamma^{j}$.
Eventually one can also add to the action a topological $B-F$ term
$\int \pi^{I}(\bar \partial A - \partial \bar A - f^{I}_{\enskip JK}A^{J}\bar 
A^{K})$. 

For our purposes we are mainly interested in the abelian case.
In particular, if the constraint, when it exists, is homogeneous, one can gauge
the scale symmetry   
\begin{equation} \label{scale}
\gamma^i \rightarrow \Lambda \gamma^i\,, ~~~~
\beta_i \rightarrow \Lambda^{-1} \beta_i,
\end{equation}
generated by the current $ J= \beta_{i}\gamma^{i}$ and
 under which  the gauge field transforms as $\bar A \rightarrow \bar A + 
\bar\partial \Lambda $.
These models describe projective spaces. 
For instance, if $\gamma^i$ 
parametrize the complex plane ${\bf C}^{n+1}$, the gauge symmetry projects 
on the projective space ${\bf CP}^n$. 

In the presence of a (homogenous) constraint the model describes a 
hypersurface in a projective space. A discussion on the subject is 
contained in the paper \cite{Grassi:2007va}.

As a first example, we consider the torus. 
It  can be viewed as a cubic hypersurface in ${\bf P}^2$ of the form
\begin{equation} \label{full}
(\gamma^1)^3 = \gamma^2 (\gamma^2 -\gamma^3)(\gamma^2 - a \gamma^3),
\end{equation}
where $a$ is the modulus of the torus. The hypersurface is homogeneous of 
degree 3
with respect to the rescaling of the fields $\gamma^i \sim \Lambda \gamma^i$.
 The gauge symmetry can be used 
to gauge to one  the coordinate $\gamma^2$  and the constraint becomes 
\begin{equation}\label{reduced}
(\gamma^1)^3 = (\gamma^3 -1)(a \gamma^3 - 1)\,,
\end{equation}
and the model can be treated as constrained in a reduced space. 

To be more explicit, following the general recipe for gauge-fixing procedure,
 we must add a single ghost field $c$ with the BRST transformations 
\begin{equation}
s\, \gamma^i = c \gamma^i\,, ~~~
s\, \beta_i = - c \beta_i\,, ~~~
s\, \bar A = - \bar \partial c\,, ~~~
s\, c =0\,, ~~~
s\, b = \rho\,, ~~~
s\, \rho =0 \,.
\end{equation}
Then, among many of possible gauge-fixing conditions, two of them are more 
interesting. 
The first one is to add to the action $S_A = \int (\beta_{i}\bar\partial
\gamma^{i} + 
\bar A \beta_{i} \gamma^{i}) $ the BRST exact term $ \int s(b \bar A)$
to reach the gauge $\bar A = 0 $. The result is the action
\begin{equation} 
S = \int (\beta_{i}\bar\partial\gamma^{i} + b \bar \partial c),
\end{equation}
with the constraint (\ref{full}) and a unbroken, rigid scale invariance.

In the second gauge-fixing, one adds to the action the BRST variation of the 
gauge fermion $ b(\gamma^2 - 1) $, that is, $\int s( b (\gamma^2 - 1))
 $ that leads to the gauge $\gamma^2 = 1 $ so that the action becomes
\begin{eqnarray}\label{gauge-fixedaction}
S = \int \Big(\beta_{i}\bar\partial\gamma^{i} + \bar A \beta_{i} \gamma^i
 + \hat \rho (\gamma^2 - 1) - b c\Big), 
\end{eqnarray}
where $\hat \rho = \rho - bc$.
Then, integrating over $ \hat \rho$, one gets the gauge-fixing condition, 
integrating over the gauge field $\bar A$, one gets that 
$\beta_2 = - \sum_{i\neq 2} \beta_i \gamma^i$ and integrating over $b$, 
one obtains $c = 0 $. The action is still free and the result is
\begin{eqnarray}\label{gauge-fixedaction2}
S = \int \Big(\sum_{i\neq 2}\beta_{i}\bar\partial\gamma^{i} \Big) \,,
\end{eqnarray}
with the constraint (\ref{reduced}). In this case as well, the rigid scale invariance is unbroken.
 

\subsection{The conifold}

The second example is the conifold constraint and the 
related Lie algebra of $D_2  = A_1 \times A_1$. This example has been already discussed 
in the introduction and it is based on the constraint (\ref{conio}). According to the $Y$-formalism, 
 the central charge is $3$ and the level of the currents $J_{ij} $ is $-1/2$.
 Moreover the OPE's of the current $J$ 
are given by
\begin{equation}
<J_{ik}(y)J(z)> = 0
\end{equation} 
\begin{eqnarray}
<T(y) J(z)> =  {{2}\over{(y-z)^{3}}} + {{1}\over{(y-z)^{2}}}J(y),
\label{qcurrent4coni}
\end{eqnarray}
\begin{eqnarray}
<J(y) J(z)> =  0.
\label{qcurrent5coni}
\end{eqnarray}
To check these results we study the model in two ways.
  
In the first way, we use a description similar to that given in \cite{Berkovits:2005hy} and 
in the second way, we use the description as a gauged linear sigma model. 

On a patch where $\gamma^1\neq 0$, we solve the conifold constraint with respect to 
the coordinate $\gamma^2 = \gamma^3\, \gamma^4/ \gamma^1$. Then, we redefine the coordinates as follows:
\begin{equation}\label{coniA}
\gamma^1 = \gamma\,,~~~~~~~
\gamma^3 = \gamma  u\,,~~~~~~~
\gamma^4 = \gamma v\,,~~~~~~~
\end{equation}
and this yields $\gamma^2 = \gamma uv$. Then, inserting these definitions in the action we can 
identify the conjugate momenta to the new fields $\gamma, u$ and $v$ by
\begin{equation}
\beta = \beta_1 + \beta_2 uv + \beta_3 u + \beta_4 v\,, ~~~~
\beta_u = \beta_2 \gamma v + \beta_3 \gamma\,, ~~~~~
\beta_v = \beta_2 \gamma u + \beta_4 \gamma\,,
\end{equation}
where $\beta, \beta_u$ and $\beta_v$ are the conjugate momenta of $\gamma, u$ and $v$, 
respectively. The new action, obtained by these field redefinitions, is again free and therefore we 
can use free OPE's for the computations. We have to notice that the redefinition of the fields is 
non-linear and the total ghost charge is carried by the field $\beta$ and $\gamma$. 
The next step is to translate the algebra of (gauge-invariant) currents $J_{ij}$ and $J$ into 
the new variables. We observe that the combinations $\beta, \beta_u$ and $\beta_v$ are gauge-invariant combinations under the gauge transformations discussed in (\ref{gaugeA}) and therefore 
the currents should be expressible in terms of those gauge-invariant basic fields. After a 
bit of algebra we get 
\begin{eqnarray}\label{coniB}
J_{12} &=& :\beta \gamma: - :\beta_u u: - :\beta_v v: \,, ~~~~~
J_{13} = :\beta_v v^2: - :\beta \gamma:v    \,, ~~~~~ \nonumber \\
J_{14} &=& :\beta_u u^2: - :\beta \gamma:u \,, ~~~~~~  
J_{23} = - \beta_u  \,, ~~~~~ \nonumber \\
J_{24} &=& - \beta_v  \,, ~~~~~ 
J_{34} = :\beta_v v: - :\beta_u u: \,, \nonumber \\
J&=& \beta \gamma\,.
\end{eqnarray}
It is easy to check that the currents $J_{ij}$ generate the $SO(4,{\mathbf C})$, but 
they have double poles with the current $J$. In order to decouple the currents $J_{ij}$ from 
the ghost currents, we add to the combinations $\beta \gamma$ an additional piece 
${1\over 2} \partial \gamma$ which can be seen as a normal ordering term. It is easy to check that the double poles generated by this new piece are 
enough to cancel the double poles between the two sets of currents and finally we can check that 
the OPE of $J$ with itself gives level zero which is consistent with $Y$-formalism (compare with 
Eq. (\ref{qcurrent5}). 

The second way of proceeding is to use the gauged linear sigma model. We 
apply this technique again to the conifold case in order to illustrate some ambiguities emerging 
from this approach (this discussion is similar to the analysis given in \cite{Grassi:2007va}). 
In order to use free coordinates plus a gauge symmetry we introduce the new fields
\begin{equation}
\label{newfields}{
\gamma^1 = a_1 a_2\,, ~~~~
\gamma^2 = a_3 a_4\,, ~~~~
\gamma^3 = a_1 a_4\,, ~~~~
\gamma^4 = a_3 a_2\,, ~~~~
} 
\end{equation}
which automatically satisfy the constraint. 
They transform as $a_i \rightarrow \Lambda a_i$ for $i=1, 3$ and 
$a_i \rightarrow a_i /   \Lambda$ for $i=2, 4$. So, we can rewrite the currents as follows:
\begin{equation}
\label{newcurrents}{
J^+_{+} = p_1 a_3\,, ~~~~
J^-_{+} = p_3 a_1\,, ~~~~
J^0_{+} = \frac{1}{2} (p_1 a_1  - p_3 a_3)\,, ~~~~
K_+ = \frac{1}{2} (p_1 a_1 + p_3 a_3), 
}
\end{equation}
and similarly by substituting $a_1, a_3$ into $a_2, a_4$ and $p_1, p_3$ into $p_2, p_4$ and 
changing the subindex $+$ into $-$. Notice that 
they form an $A_1 \times GL(1) \times A_1 \times GL(1)$ algebra. 

Expressed in terms  of the variables $\beta_{i}$ and $\gamma^{j}$ the currents
$J_{\pm}^{\pm}$, $J_{\pm}^{0}$ are linear combinations of $J_{ij}$ and 
\begin{eqnarray}
K_{+} = K_{-}= \frac{1}{2} J,
\end{eqnarray}
so that 
\begin{eqnarray}
\hat J = K_{+} - K_{-},
\end{eqnarray}
vanishes. Then in
 the model expressed in term of the variables $p_{i}$ and $a^{i}$, $\hat J$
is a constraint and the model is a gauged model with action
\begin{eqnarray}
S = \int ( p_{i} a^{i} + \bar A \hat J ).
\end{eqnarray}
$\hat J$ is a primary field with vanishing anomaly (i.e., vanishing triple pole
in its OPE with $T$) and   
\begin{equation}
\label{ghostcurrent1}
{<\hat J(z) \hat J(w)> = 0\,.}
\end{equation}
The fact that there is no double pole in the ghost current can be 
checked by using the Nekrasov method 
\cite{Nekrasov:2005wg} 
starting from the character of the zero modes $\chi(t) = (1- t^2)/(1-t)^4$ 
by substituting $t \rightarrow  e^x$ and expanding the result 
as a polynomial of $x$ and $\log(x)$. Taking into account of the contribuion 
of the ghost fields $b$, $c$ coming from the gauge fixing, the central charge 
is $2$. As for the gauge current 
\begin{eqnarray}
J \equiv K_{+} + K_{-},
\end{eqnarray}
one has 
\begin{eqnarray}
<T(z) J(w)> &=& \frac{2}{(z-w)^3} + {{J(z)} \over {(z-w)^2}},
\nonumber\\
<J(z) J(w)> &=&  - \frac{1}{(z-w)^2}.
\end{eqnarray}
Moreover the level of the OPE among the currents that generate 
$(A_1) \times (A_1)$ turns out to be $-1/2$. These features agree
with the central charge, anomaly and levels computed using the $Y$-formalism
(as well as using the techniques in \cite{Berkovits:2005hy}) except for the 
OPE   $ <J(z) J(w)>$ which vanishes in the $Y$-formalism computation.
However notice that, due  to the constraint, there is an ambiguity in the definition 
of the current $J$, in the model expressed in terms of $p_{i}$ and $a^{i}$.
Indeed we can always add a term proportional to the constraint 
\begin{equation}
J = (K_+ + K_-)  + \alpha (K_+ - K_-),
\end{equation}
 which changes the double poles (they are contact terms) but does not 
 change the rest of the algebra.
 It is easy to check that the current $J$ has zero level if we set 
$\alpha = \pm i$. 
It should be possible to decompose the 
character formula given in \cite{Grassi:2007va} 
in terms of the corresponding characters. Notice that all characters have 
negative levels and therefore we expect some singular vectors of the Verma
 modules of the corresponding field theories.  

\section{Adding other variables}

In general the models described above are purely bosonic and are lacking in the
 BRST charge to construct the physical space of states. Moreover they  fail 
to provide a conformal field theory with zero central charge. In order to 
overcome these problems, we can add new variables. The natural setting is to 
reproduce the pure spinor construction,  adding some 
bosonic variable $p, X$ and some fermionic variables $p_i, \theta^i$ with 
$p(z) X(w) \rightarrow (z-w)^{-1}$ and $p_i(z) \theta^j(w) \rightarrow 
\delta^j_i (z-w)^{-1}$.
 We assume that 
there is only a constraint $\Phi(\gamma) = 0$ . The 
index for the fermionic variables is the same as that for the bosonic
 non-linear sigma model ones, described by $\gamma^i$. 

The setting is described in paper \cite{Grassi:2005jz}, where it has been 
shown the structure of the model, 
and the gauge-invariant operators. In \cite{Grassi:2005jz}, a simple 
example of constraint $\Phi(\gamma) = \gamma^1 \gamma^2$ has been taken 
into account, but the technique can be used for more general examples 
of the form 
$\Phi(\gamma) = 1/2 \gamma^i g_{ij} \gamma^j$. 
In addition to the  operators $J$ and $N_{i j}\equiv J_{ij}$, 
which are gauge invariant under the gauge 
symmetry $\beta_i = \Lambda g_{ij} \gamma^j$, we shall consider the operators
 \begin{equation}\label{defs}
\Pi = p + {1\over 2} \theta^i g_{ij} \partial \theta^j\,,   ~~~~
d_i = p_i + {1\over 2} \partial X g_{ij} \theta^j,
\end{equation}
so that
\begin{eqnarray}
<\Pi(z) \Pi(w)> &=& 0, \nonumber\\
<d_{i}(z) \Pi(w)> &=&  {{1}\over{z-w}} \partial\theta_{i},
\nonumber\\
<d_i(z)d_j(w)> &=& {{1}\over{z-w}} g_{i j}\partial X.
\end{eqnarray}
Then one can establish a nilpotent BRST charge in the same way as in the pure spinor formalism
\begin{equation}\label{BRST}
Q = \oint \gamma^i d_i.
\end{equation}
Notice that the presence of the constraint $\Phi(\gamma)=0$ is essential to 
have a nilpotent BRST charge. Using this 
charge one can construct the physical space of states.   

As is known, the gauge symmetry (\ref{gaugesymm}) implies that the physical 
quantities 
should be gauge invariant under it. This implies that only the gauge-invariant combinations containing the field $\beta_i$ 
are  $ N_{ij} \equiv J_{ij}$, $J$ and $T$ of zero ghost number as 
defined in (\ref{current}), (\ref{currentbis}) and (\ref{currentter}), respectively. Therefore, there do not exist $Y$-independent operators with
negative ghost number which permit the construction 
of an operator $B$ with ghost number $-1$ such that 
\begin{equation} \label{brst-b}
[Q, B] = \hat T.
\end{equation}
Here $\hat T $ is the total energy momentum tensor of the model given by
\begin{eqnarray}
\hat T = p \partial X + p_i \partial \theta^i  + T,
\end{eqnarray}
where $T$ is the energy 
momentum tensor of the $\beta-\gamma$ system given by (\ref{currentter}). 
However, as has been noticed in  \cite{Berkovits:2004px} for the case of pure spinors, one 
can construct a sequence of 
operators $\{G^i, H^{[ij]}, \cdots\}$  
that satisfy the following equations 
\footnote{ $:\gamma^i \hat T :$  has the normal ordering term $ -1/2 
\partial^{2} \gamma^{i}$ (to be primary).}  
\begin{equation} \label{sequence}
[Q, G^i] = :\gamma^i \hat T :\,, ~~~~
[Q, H^{[ij]}] = \gamma^{[i} G^{j]}\,, ~~~~  \cdots
\end{equation}
where
\begin{equation}\label{bfield}
G^i = \Pi d^i +2 N^{i j}\partial \theta_{j} \,, ~~~~
H^{[ij]} = N^{i j}\Pi  \,,~~~~
\gamma^{[i} H^{jk]} =0.
\end{equation}
It is intersting to observe that also in this simple model 
this construction can be done and looks simpler than in the pure spinor case 
since here there is no gamma matrices structure to take into considerations. 
Equations (\ref{brst-b}) can be checked immediatly at the classical level,
but we have verified that they hold also at the quantum level: indeed the 
$Y$-dependent terms that arise in (\ref{bfield}) combine in an appropriate way
to give (\ref{brst-b}) if one takes in account correctly the subtleties 
of the normal reordering through the so-called rearrangement theorem (see 
\cite{Y-formalism:2007ab} for details of the procedure).  

With the aid of $G^{i}$ one can construct a $Y$-dependent $B$-ghost 
\begin{eqnarray}
B^{(Y)} = Y_{i}G^{i},
\end{eqnarray}
which satisfies (\ref{brst-b}). However, with the help of both
$G^{i}$ and $H^{ij}$ one can  construct a $Y$-independent  $B$-field, by 
the so-called minimal approach, introducing the picture-changing operators 
$Z_B = [Q, \Theta(B J + B^{ij} N_{ij})]$ and $Y_C = C_i \theta^i \delta(C_j 
\gamma^j)$, and 
constructing the field $B_B$ such that 
\begin{eqnarray}
[Q, B_B] = Z_B T.
\end{eqnarray}
Moreover it is also possible to imitate the non-minimal approach of 
\cite{Berkovits}  by introducing $N$ quartets of new additional fields 
and get a consistent expression for $B$ (as opposed to $B_B$) in terms of the 
operators (\ref {bfield}).

The bosonic new fields are $\lambda_{i}$, with momenta $w^{i}$ and the fermonic
ones are $r_{i}$ with momenta $s^{i}$ with the constraints 
\begin{equation}
 \lambda_{i}g^{ij}\lambda_{j}=0 \, ~~~~ \lambda_{i}g^{ij}r_{j}=0.
\end{equation} 
 The new BRST charge is 
$ Q = \oint \gamma^i d_i + \oint w^{i}r_{i}$ and of course the new $\hat T$
contains two new terms coming from these quartets.Also
the $Y$-formalism can be 
extended to this sector following \cite{Y-formalism:2007ab}.

Then the $B$-field is
\footnote{ $:{{\lambda_{i}}\over{(\gamma\lambda)}}G^{i}:$ contains normal 
ordering terms that we will not specify.}
\begin{equation}
 B = s^{i}\partial \lambda_{i} + :{{\lambda_{i}}\over{(\gamma\lambda)}}G^{i}:
 -2{{\lambda_{i} r_{j}}\over {(\gamma\lambda)^{2}}}H^{ij},
\end{equation}
and satisfies (\ref{brst-b}).
One can verifies that, as in the pure spinor case, this $B$-field is 
cohomological equivalent to the $Y$-dependent $B$-field, $B^{(Y)}$.  

\section{Superprojective spaces}
\subsection{An example: the supercone constraint}

We would like to provide an example of conformal field theory 
where the constraint (which emerges by requiring that the BRST charge is 
nilpotent) is
\begin{eqnarray}
x y + i \theta^1 \theta^2 = 0,
\label{coneA}
\end{eqnarray}
where $x,y$ are bosonic coordinates and $\theta^i$ are fermionic coordinates. 
This constraint defines a supercone.
To derive the constraint (\ref{coneA}) we can start from the supergroup $Osp(2|2)$. 
This supergroup is described in terms of 4 bosonic generators $H, \widetilde H, E^\pm$ and 
4 fermionic ones $Q_\alpha, Q_{\hat \alpha}$ with $\alpha, \hat \alpha =1,2$. 
The algebra $osp(2|2)$ is easily described by the (anti)commutators 
\begin{eqnarray}\label{osp}
&&[H, E_\pm] = 
\pm 2E_\pm\,, ~~~~~
[E^+ , E^-] = H\,, ~~~~~
[H, \widetilde H] = 0\,,  ~~~~~~ \nonumber \\
&&[\widetilde H, E_\pm] = 0\,, ~~~~~~~
[\widetilde H, Q_\alpha ] = \epsilon_{\alpha\beta} Q_{\beta} \,, ~~~~~
[\widetilde H, Q_{\hat\alpha} ] = \epsilon_{\hat\alpha\hat\beta} Q_{\hat\beta} \,, ~~~~~ 
\nonumber \\
&&[H, Q_\alpha ] = Q_\alpha\,,~~~~~
[H, Q_{\hat\alpha} ] = - Q_{\hat\alpha}\,,~~~~~ \\
&&
\{Q_\alpha, Q_\beta \} = {1\over2} \delta_{\alpha \beta} E^+ \,,~~~~~ 
\{Q_{\hat \alpha}, Q_{\hat\beta} \} = {1\over2} \delta_{\hat\alpha \hat\beta} E^- \,,~~~~~ 
\{Q_{\alpha}, Q _{\hat \alpha} \} = {1\over 2} \delta_{\alpha \hat\alpha} H + {1\over 2} \epsilon_{\alpha\hat\alpha} \widetilde H\,, ~~~~ \nonumber \\
&& [E^+, Q_{\alpha}] = [E^-, Q_{\hat \alpha}] = 0\,, ~~~~  
 [E^+, Q_{\alpha}] = -\delta_{\alpha\hat\alpha} Q_{\alpha} \,, ~~~~  
 [E^+, Q_{\hat\alpha}] = -\delta_{\alpha \hat\alpha} Q_{\hat \alpha}\,.\nonumber  
\end{eqnarray}

We define a BRST charge of the form
\begin{equation}\label{newq}
Q = \oint \Big( c^+ E_+ +   c^- E_- + \lambda^{\alpha} Q_{\alpha} 
+ \lambda^{\hat\alpha} Q_{\hat\alpha} + {\rm bcc ~~ terms} \Big),
\end{equation}
where we have "gauged" the generators of the coset $E_\pm, Q_\alpha, Q_{\hat\alpha}$. 
By computing the nilpotency of the BRST charge (for simplicity we have 
computed only the simple poles) one gets that 
\begin{equation}\label{newpA}
\{Q, Q\} = \oint \Big( c^+ c^- + {1\over 2} \lambda^{\alpha} \delta_{\alpha \hat \alpha} \bar\lambda^{\hat \alpha} \Big)  H + \Big({1\over 2} 
\lambda^{\alpha} \epsilon_{\alpha \hat \alpha} \bar\lambda^{\hat \alpha} \Big)  \widetilde H,
\end{equation}
 where the non-nilpotentcy terms are proportional to the 
 Cartan generators $H$ and $\widetilde H$. So, in order that the entire BRST charge is nilpotent 
 we need that 
 \begin{equation}
\Big( c^+ c^- + {1\over 2} \lambda^{\alpha} \delta_{\alpha \hat \alpha} \bar\lambda^{\hat \alpha} \Big) =0\,, 
~~~~~
\Big(\lambda^{\alpha} \epsilon_{\alpha \hat \alpha} \bar\lambda^{\hat \alpha} \Big)  =0\,.
\end{equation}
To solve the second one, we can set $\bar\lambda^{\hat\alpha} = \lambda^{\alpha}$ and 
the first one becomes the supercone constraint (\ref{coneA}) after a change of coordinates $x = \lambda^1 + i \lambda^2$ and $y = \lambda^1 - i \lambda^2$. 

A detailed discussion on the interpretation of the BRST charge (\ref{newq}) can be found in 
\cite{Grassi:2004cz}. The cohomology, the interpretation of the present model, is completely 
open and it deserves some study. 

\subsection{$Y$-formalism for superprojective spaces} \par

One of motivations behind this study is 
that this new model might be a prototype for supertwistors where not only
very limited class of amplitudes are evaluated but also quantum algebra is still unknown. We construct $Y$-formalism for this class of models in
this section and wish to clarify these problems in the future work.

The constraint to which we turn our attention takes the form of a
super-cone (\ref{coneA}). For simplicity and generalization to other cases,
it is convenient to rewrite the super-cone constraint (\ref{coneA}) as
\begin{eqnarray}
\Phi(\gamma^i, \theta^a) \equiv \gamma^i g_{ij} \gamma^j 
+ \theta^a h_{ab} \theta^b = 0,
\label{constraintA}
\end{eqnarray}
where $\gamma^{i}$ ($\theta^{a}$) are commuting (anticommuting) fields and 
$i = 1, \cdots, 2N_{1}$, $a = 1, \cdots. 2N_{2}$ and
\begin{eqnarray}
g_{i,N_{1}+i} &=& g_{N_{1}+i, i} = \frac{1}{2}, \nonumber\\
h_{i,N_{2}+ i} &=& - h_{N_{2}+i,i} = \frac{1}{2} i,
\label{metric}
\end{eqnarray}
and otherwise are vanishing. We shall present the calculations only for the case
\begin{eqnarray}
N_{1} = N_{2} = N.
\end{eqnarray}
The extension to the general case is straightforward.  

Let us start with the action:
\begin{eqnarray}
S = \int ( \beta_i \bar \partial \gamma^i - p_a \bar \partial \theta^a).
\label{actionA}
\end{eqnarray}
This action is invariant under the gauge transformations
\begin{eqnarray}
\delta \beta_i  &=& \lambda \gamma_i,  \nonumber\\
\delta p_a &=& - \lambda \theta_a,
\label{gaugeA1}
\end{eqnarray}
where we have defined the variables with lower indices as
\begin{eqnarray}
\gamma_i &=& \gamma^j g_{ji},  \nonumber\\
\theta_a &=& \theta^b h_{ba}.
\label{index}
\end{eqnarray}

Now let us construct classically gauge-invariant currents whose expressions
are given as follows:
\\
a) \underline{Ghost current} \par
\begin{eqnarray}
J^0 = \beta_i \gamma^i - p_a \theta^a.
\label{ghost current}
\end{eqnarray}
b) \underline{Stress-energy tensor} \par
\begin{eqnarray}
T^0 = \beta_i \partial \gamma^i - p_a \partial \theta^a.
\label{stress}
\end{eqnarray}
c) \underline{'Lorentz' current} \par
\begin{eqnarray}
J_{ij}^0 = \beta_{[i} \gamma_{j]}.
\label{Lorentz}
\end{eqnarray}
d) \underline{'Lorentz' current in superspace} \par
\begin{eqnarray}
j_{ab}^0 = p_{(a} \theta_{b)}.
\label{superLorentz}
\end{eqnarray}

Next, let us construct the $Y$-formalism. Firstly, let us introduce
two kinds of $Y$-fields by
\begin{eqnarray}
Y_i &=& \frac{v_i}{v_k \gamma^k + w_c \theta^c}, \nonumber\\
Y_a &=& \frac{w_a}{v_k \gamma^k + w_c \theta^c}.
\label{Y}
\end{eqnarray}
The constant vectors $v_i$ and $w_a$ are commuting and anticommuting numbers, respecively. 
\footnote{For simplicity, we have used the additional constraints
$v_i v^i + w_a w^a = 0$ or equivalently, in the $Y$-fields, we have 
$Y_i Y^i + Y_a Y^a = 0$. They are not essential, but they simplify the computations. 
}
Moreover, the above definition of $Y$-fields leads to the following
useful relations:
\begin{eqnarray}
1 &=& Y_i \gamma^i + Y_a \theta^a, 
\nonumber\\
\partial Y_i &=& - (Y_k \partial \gamma^k + Y_c \partial \theta^c) Y_i,
\nonumber\\
\partial Y_a &=& - (Y_k \partial \gamma^k + Y_c \partial \theta^c) Y_a,
\label{relations}
\end{eqnarray}
which will be often utilized in calculating an algebra below.

Secondly, we make the following projection operators:
\begin{eqnarray}
K_i \ ^j &=& Y^j \gamma_i = \frac{v^j \gamma_i}{v_k \gamma^k + w_c \theta^c}, \nonumber\\
K_i \ ^a &=& Y^a \gamma_i = \frac{w^a \gamma_i}{v_k \gamma^k + w_c \theta^c}, 
\nonumber\\
\nonumber\\
K_a \ ^b &=& Y^b \theta_a = \frac{w^b \theta_a}{v_k \gamma^k + w_c \theta^c},
\nonumber\\
K_a \ ^i &=& Y^i \theta_a = \frac{v^i \theta_a}{v_k \gamma^k + w_c \theta^c}. 
\label{projection}
\end{eqnarray}
Using these definition of projection operators together with the super-cone
constraint and the relations (\ref{relations}), we can derive the 
following useful equations: 
\begin{eqnarray}
K_i \ ^i - K_a \ ^a &=& K_i \ ^j K_j \ ^i - K_a \ ^b K_b \ ^a
- 2 K_i \ ^a K_a \ ^i  = 1, 
\nonumber\\
K_j \ ^i \gamma^j + K_b \ ^i \theta^b &=& K_j \ ^a \gamma^j + K_b \ ^a \theta^b = 0, 
\nonumber\\
K_j \ ^i \partial \gamma^j + K_b \ ^i \partial \theta^b 
&=& K_j \ ^a \partial \gamma^j + K_b \ ^a \partial \theta^b = 0, 
\nonumber\\
Y_j K_i \ ^j  + Y_b K_i \ ^b &=& Y_j K_a \ ^j + Y_b K_a \ ^b = 0, 
\nonumber\\
Y_j \partial K_i \ ^j  + Y_b \partial K_i \ ^b 
&=& Y_j \partial K_a \ ^j + Y_b \partial K_a \ ^b = 0, 
\label{relations2}
\end{eqnarray}
which will be also used in calculating an algebra.

Thirdly, we set up the basic OPE's
\begin{eqnarray}
< \beta_i(y) \gamma^j(z) > &=& \frac{1}{y-z} (\delta_i^j - K_i \ ^j(z)), \nonumber\\
< \beta_i(y) \theta^a(z) > &=& - \frac{1}{y-z} K_i \ ^a(z), 
\nonumber\\
< p_a(y) \theta^b(z) > &=& \frac{1}{y-z} (\delta_a^b - K_a \ ^b(z)),
\nonumber\\
< p_a(y) \gamma^i(z) > &=& \frac{1}{y-z} K_a \ ^i(z).
\label{OPE}
\end{eqnarray}
Then, it is easy to show that $< \beta_i(y) \Phi(z) > = < p_a(y) \Phi(z) >
= 0$.

Now we move on to a derivation of an algebra among gauge-invariant quantum currents.
Requiring that terms depending on $Y$-fields, which violate the Lorentz symmetry,
should be absent in the algebra, we can fix the expression of gauge-invariant quantum currents uniquely. Actually, the currents read
\\
a) \underline{Ghost current} \par
\begin{eqnarray}
J = \beta_i \gamma^i - p_a \theta^a + \frac{3}{2} (Y_i \partial \gamma^i
+ Y_a \partial \theta^a).
\label{q-ghost current}
\end{eqnarray}
b) \underline{Stress-energy tensor} \par
\begin{eqnarray}
T = \beta_i \partial \gamma^i - p_a \partial \theta^a 
+ \frac{1}{2} \partial (Y_i \partial \gamma^i + Y_a \partial \theta^a).
\label{q-stress}
\end{eqnarray}
c) \underline{'Lorentz' current} \par
\begin{eqnarray}
J_{ij} = \beta_{[i} \gamma_{j]} - Y_{[i} \partial \gamma_{j]}
- \frac{1}{2} \partial Y_{[i} \gamma_{j]}.
\label{q-Lorentz}
\end{eqnarray}
d) \underline{'Lorentz' current in superspace} \par
\begin{eqnarray}
j_{ab} = p_{(a} \theta_{b)} + Y_{(a} \partial \theta_{b)}
+ \frac{1}{2} \partial Y_{(a} \theta_{b)}.
\label{q-superLorentz}
\end{eqnarray}

It then turns out that the algebra among currents is of form
\begin{eqnarray}
<J_{ij}(y) J_{kl}(z)> &=& \frac{1}{2 (y-z)^2} 
(g_{k[i} g_{j]l} - g_{l[i} g_{j]k})  
+ \frac{1}{y-z} (g_{k[i} J_{j]l} - g_{l[i} J_{j]k}),
\nonumber\\
<j_{ab}(y) j_{cd}(z)> &=& \frac{1}{2 (y-z)^2} 
(h_{c(a} h_{b)d} + h_{d(a} h_{b)c})  
+ \frac{1}{y-z} (h_{c(a} j_{b)d} + h_{d(a} j_{b)c}),
\nonumber\\
<J(y) J_{ij}(z)> &=& 0,
\nonumber\\
<J(y) j_{ab}(z)> &=& 0,
\nonumber\\
<T(y) J_{ij}(z)> &=& \frac{1}{(y-z)^2} J_{ij}(y),
\nonumber\\
<T(y) j_{ab}(z)> &=& \frac{1}{(y-z)^2} j_{ab}(y),
\nonumber\\
<T(y) T(z)> &=& \frac{-1}{(y-z)^4} + \frac{1}{(y-z)^2}
(T(y) + T(z)),
\nonumber\\
<T(y) J(z)> &=& \frac{-2}{(y-z)^3} + \frac{1}{(y-z)^2} J(y),
\nonumber\\
<J(y) J(z)> &=& \frac{4}{(y-z)^2}. 
\label{algebra2}
\end{eqnarray}
Compared this algebra with that of the bosonic $\beta-\gamma$ system, we find that this system corresponds to the case $N = 0$. This fact can be easily
understood since the dynamical degrees of freedom are canceled between
bosons and fermions.

\section*{Acknowledgments} 
The work of P.A.G. and of M.T. is supported in part by the European Community's Human Potential Programme under contract MRTN-CT-2004-005104, "Constituents, Fundamental Forces and Symmetries of the Universe". The work of M.T. is also supported  
by INTAS 05-10000087928. P.A.G. would like to thank C. Imbimbo, S. Giusto and G. Policastro for useful discussions. I.O. is grateful to T. Tokunaga for
valuable discussions.


 \end{document}